 \documentclass[preprint,showpacs,preprintnumbers,amsmath,amssymb,pre]{revtex4}

\usepackage{graphicx}
\usepackage{dcolumn}
\usepackage{bm}
\usepackage[latin1]{inputenc}
\usepackage[T1]{fontenc}
\usepackage{ae,aecompl}

\begin{document}

\newcommand{\rar}{$\rightarrow$}
\newcommand{\lrar}{$\leftrightarrow$}
\newcommand{\beq}{\begin{equation}}
\newcommand{\eeq}{\end{equation}}
\newcommand{\bea}{\begin{eqnarray}}
\newcommand{\eea}{\end{eqnarray}}
\newcommand{\Req}[1]{Eq.\ (\ref{E#1})}
\newcommand{\req}[1]{(\ref{E#1})}
\newcommand{\degree}{$^{\rm\circ} $}
\newcommand{\pcite}{\protect\cite}
\newcommand{\pref}{\protect\ref}
\newcommand{\Rfg}[1]{Fig.\ \ref{F#1}}
\newcommand{\rfg}[1]{\ref{F#1}}
\newcommand{\Rtb}[1]{Table \ref{T#1}}
\newcommand{\rtb}[1]{\ref{T#1}}

\title{Analysis of Accordion DNA Stretching Revealed by The Gold Cluster
Ruler}

\author{Alexey K. Mazur}
\email{alexey@ibpc.fr}
\affiliation{CNRS UPR9080, Institut de Biologie Physico-Chimique,
13, rue Pierre et Marie Curie, Paris,75005, France.}


\begin{abstract}
A promising new method for measuring intramolecular distances in
solution uses small-angle X-ray scattering interference between gold
nanocrystal labels (Mathew-Fenn et al, Science, {\bf322}, 446 (2008)).
When applied to double stranded DNA, it revealed that the DNA length
fluctuations are strikingly strong and correlated over at least 80
base pair steps. In other words, the DNA behaves as accordion bellows,
with distant fragments stretching and shrinking concertedly. This
hypothesis, however, disagrees with earlier experimental and
computational observations. This Letter shows that the discrepancy can
be rationalized by taking into account the cluster exclusion volume
and assuming a moderate long-range repulsion between them. The
long-range interaction can originate from an ion exclusion effect and
cluster polarization in close proximity to the DNA surface.
\end{abstract}

\pacs{87.14.gk 87.15.H- 87.15.ap 87.15.ak}

\maketitle

Long double stranded DNA (dsDNA) behaves as a continuous elastic rod
placed in a heat bath (Ref. \onlinecite{Cantor:80a}, Ch. 19). The
fluctuations of the length, $L$, and the end-to-end distance, $R$, are
characterized by the the canonical variances $\sigma_R^2(L)$ and
$\sigma_L^2(L)$, respectively. Function $\sigma_L^2(L)$ should be
linear because the stacking interactions are short-range and, starting
from a few base pair steps (bps), the dsDNA can be considered as a
concatenation of short independent fragments. The small-angle X-ray
scattering interference (SAXSI) between gold nanocrystal labels makes
possible very accurate measurements of short intramolecular distances
\cite{Mathew-Fenn:08a}. When the length of dsDNA was measured it
turned out that function $\sigma_L^2(L)$ is approximately quadratic
suggesting that the length fluctuations are positively correlated over
at least 80 bps \cite{Mathew-Fenn:08b}, that is, dsDNA breathes as
accordion bellows, with distant fragments stretching and shrinking
concertedly. This conclusion questions the current physical models of
the double helix and offers a simple possibility of long-range
communications along DNA, with broad mechanistic implications for gene
regulation. It was argued, however \cite{Mzlanl:09}, that the SAXSI
data are not entirely consistent with earlier observations. Although
the $\sigma_L^2(L)$ profile was not accurately checked before, the
amplitudes of stretching fluctuations measured by other methods differ
from the SAXSI results.  Notably, the amplitudes of fluctuations
observed in the atom force microscopy experiments
\cite{Rivetti:01,Sanchez-Sevilla:02,Podesta:04} were much smaller than
should be expected according to the SAXSI data \cite{Mzlanl:09}. The
effect discovered by Mathew-Fenn et al.  \cite{Mathew-Fenn:08b}
probably has a different physical origin and in the present Letter we
propose an alternative interpretation of these results. We show that
the striking quadratic growth of $\sigma_L^2(L)$ can be caused by the
cluster-DNA excluded volume effect and weak cluster-cluster repulsion.
The long-range effective repulsion between the gold labels probably
results from the solvent polarization induced by the strong
electrostatic field near the DNA surface.

\begin{figure}[htb]
 \centerline{\includegraphics[width=6cm]{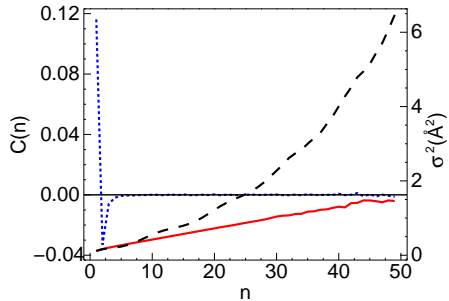}}
\caption{\label{Fmddd} Color online.
Stretching fluctuations in dsDNA according to a 208 ns of all-atom MD
simulation. The left y-axis and the blue dotted line show the behavior
of the correlation terms $C(n)$. Variances $\sigma_L^2(n)$ and
$\sigma_R^2(n)$ are shown by the solid red and dashed black lines,
respectively, with the values on the right y-axis. The
details of MD protocols can be found elsewhere \cite{Note1}.}
\end{figure}

We begin with analysis of stretching fluctuations in realistic
all-atom MD simulations of a 50-mer fragment of a poly-GC dsDNA
(\Rfg{mddd}). The variance of the MD fluctuations of the DNA length
can be expanded as \cite{Note1}
$$
\sigma_L^2(n)=nC(1)+\sum_{m=2}^nC(m),
$$
where $n$ is the chain length. Terms $C(m)$ include correlations of
the helical rise in di-nucleotides separated by $(m-1)$ steps. Term
$C(1)$ is positive-definite and function $\sigma_L^2(n)$ is strictly
linear when single step fluctuations are uncorrelated along the chain
($C(m)=0$, m>1). \Rfg{mddd} reveals that stretching fluctuations at
neighboring steps are anticorrelated, but the correlations fade away
over about four bps.  The $C(2)$<0 gives a small positive y-intercept
of the $\sigma_L^2(n)$ profile, and this is the only distinguishable
deviation from strict linearity. The bending persistence length,
$l_b$, and the stretching Young modulus, $Y_f$, measured as described
elsewhere \cite{Mzbj:06,Mzjpc:08,Mzjpc:09} are about 100 nm and 4700 pN,
respectively, indicating that the MD forcefield overestimates the
rigidity of dsDNA \cite{Note2}. To correct for this discrepancy the
upper estimate of the variances in \Rfg{mddd} should be increased by a
factor of three, and yet this gives an order of magnitude smaller
values than in experiment \cite{Mathew-Fenn:08b}.  In the course of
dynamics, the $C(m)$ terms with $m$>4 gradually decrease with time,
with the linearity of the $\sigma_L^2(n)$ plot improved. Therefore,
the main features of the pattern in \Rfg{mddd} hardly depend upon the
limited MD sampling. Similar linear $\sigma_L^2(n)$ profiles were also
obtained for shorter GC- and AT-alternating dsDNA
\cite{Mzbj:06,Mzjpc:09}.

Our MD simulations are not meant to disprove experimental
observations, but they demonstrate that the most detailed dsDNA models
currently used do not involve interactions that might cause long-range
stretching correlations when bound to large gold nanoparticles. At
the same time, the $\sigma^2_R(L)$ plot in \Rfg{mddd} has a quadratic
profile, and one can conjecture that gold cluster SAXSI measurements
suffer from incomplete subtraction of the bending contribution from
the measured $\sigma^2_R$. Indeed, the original interpretation
\cite{Mathew-Fenn:08b} assumed that the labels make a constant
contribution to $\sigma_R^2$, which would be true if they were freely
rotating around the points of attachment. In reality, these are bulky
objects that, due to the excluded volume effect, on average elongate
DNA, and their contribution to $\sigma^2_R(L)$ should vary with $L$.
The clusters are restrained by short as well as long-range
interactions and below we try to figure out under which conditions
this effect can explain the gold cluster SAXSI data.

Different hypotheses were examined by using BD simulations with
earlier tested discrete WLC models \cite{Mzjpc:08}. The number of
beads, the link length, and the harmonic potentials were adjusted to
model dsDNA fragments with $l_b=50$ nm, and $Y_f$=1000 pN.  The BD
results are compared with experimental data for different intact
double helices \cite{Mathew-Fenn:08b} and a hybrid construct from Ref.
\onlinecite{Mathew-Fenn:08a} schematically shown in \Rfg{plosdd}. This
construct consisted of two 12-mer dsDNA linked by a flexible hinge of
three unpaired thymine bases. The poly-thymine single stranded DNA
(ssDNA) has a negligible bending stiffness \cite{Goddard:00}. The gold
clusters were modeled as terminal beads attached by 1.5 nm flexible
links with the average bend deviation angle $\bar\theta=46^\circ$
obtained by fitting to the experimental $\sigma^2_R$ for 35 bp dsDNA
\cite{Mathew-Fenn:08b}. This corresponds to relatively free particles,
but they never point backward with respect to the chain direction. The
single stranded DNA (ssDNA) was modeled by three beads joined by 0.5
nm links. The best agreement with the experimental data
\cite{Mathew-Fenn:08a} was observed with the stiffness of the ssDNA
link corresponding to a very small $l_b$ value around 0.5 nm.

\begin{figure}[htb]
 \centerline{\includegraphics[width=6cm]{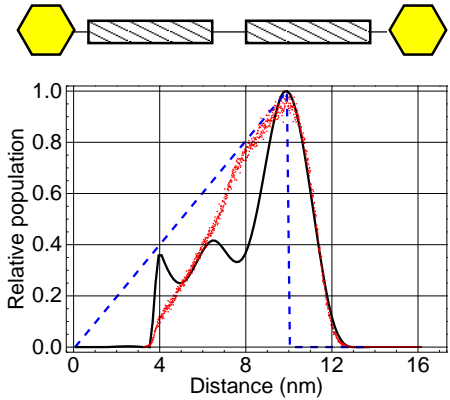}}
\caption{\label{Fplosdd} Color online. End-to-end distance
distributions for the construct sketched on top. The gold nanocrystal
labels and the dsDNA fragments are shown by hexagons and dashed
rectangles, respectively. The distribution shown by the solid black
line is constructed from the experimental data \cite{Mathew-Fenn:08a}
by using cubic spline interpolation between several key points. The
ideal theoretical distribution is shown by the dashed blue line. The
dotted red trace features the results of BD simulations.}\end{figure}

The effective diameter of a passivated gold cluster significantly
exceeds that of its core. This is clearly seen from the data for the
hybrid construct (see \Rfg{plosdd}).  Approximated as two thin rigid
sticks joined by a hinge, it should give a triangular end-to-end
distance distribution. The experimental distribution vanishes below
3.5 nm, which can only be due to the cluster-cluster exclusion.  The
contact radius of about 1.7 nm can be obtained as the radius of the
nanocrystal core (0.7 nm) plus the thioglucose passivating shell (0.6
nm) plus a layer of hydration water (0.4 nm).  The experimental
profile in \Rfg{plosdd} features two local energy minima on the
effective interaction potential, however, it is repulsive on average
and apparently long-range. The figure also shows the results of BD
simulations of the excluded volume effect.  The short-range
non-neighbor interactions were modeled by a flat-bottom harmonic
repulsion starting from the contact distances corresponding to DNA and
cluster radii of 1 and 2 nm, respectively.  The distance distribution
follows the analytical form for large separations and progressively
becomes depleted for distances below 8 nm where the experimental
distribution exhibits a complex structure.

\begin{figure}[htb]
 \centerline{\includegraphics[width=8.6cm]{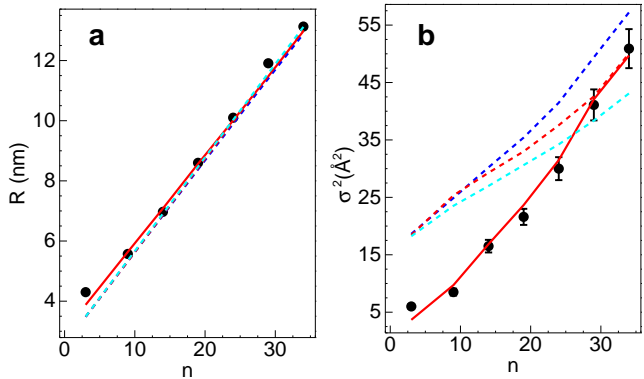}}
\caption{\label{Fscidd} Color online. Comparison of experimental
and computed cluster-cluster distances (a) and
distance variances (b). Black circles with error bars show the
experimental points published in the supplement to Ref.
\onlinecite{Mathew-Fenn:08b}. The dashed lines show BD results for
$l_b$=40 nm (upper blue), 50 nm (middle red), and 60 nm (lower cyan),
without cluster-cluster interactions. The solid red lines correspond
to $l_b$=50 nm with the fitted interaction potential.} \end{figure}

The experimental data for labels attached to intact dsDNA
\cite{Mathew-Fenn:08b} are analyzed in \Rfg{scidd}. \Rfg{scidd}(a)
evidences that the 1.5 nm link used for attachment of the labels
provides satisfactory agreement of computed and experimental DNA
lengths regardless of the varied parameters. This value approximately
equals the real length.  Comparison of the three dashed traces in
\Rfg{scidd}(b) demonstrates that the label mobility and the DNA
flexibility produce a synergetic effect and increase the bending
contribution to the measured $\sigma_R^2$. The bending part of the
variance for 35 bp DNA with $l_b$=50 nm is about 0.07 nm$^2$, which is
less than the increment obtained by a 10 nm variation of $l_b$.

For short DNA the three dashed traces converge to the same variance
corresponding to the distinct contribution of the cluster mobility.
This value is larger than in experiment. However, in the shortest DNA
fragments the labels should be significantly restrained because they
almost touch one another (see \Rfg{plosdd}). The effective
cluster-cluster potential can be reconstructed from the experimental
$\sigma_R^2(n)$ dependence in \Rfg{scidd}(b). One can reasonably assume
that these interactions cease at about 10 nm. We took this point as
zero, used some trial positive energy for 8 nm, and assumed that this
increment doubles with the distance reduced by every 2 nm.  Between
these reference points the cubic spline interpolation was applied.
Thus defined potential was introduced into the BD simulation algorithm and
fitted by using the experimental variances. As seen in \Rfg{scidd}, the
resulting discrete WLC model accurately reproduces the experimental
observations.

\begin{figure}[htb]
 \centerline{\includegraphics[width=8.6cm]{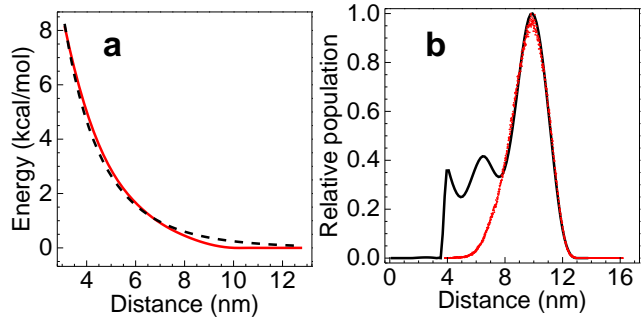}}
\caption{\label{Fpotfit} Color online. (a) The cluster-cluster
interaction potential fitted to experimental data (solid red line)
compared with Debye-H\"uckel potential with parameters given in the
text (dashed black line). (b) Comparison of the experimental
end-to-end distance distributions for the model system shown in
\Rfg{plosdd} (solid black line) with the results of BD simulations
with the fitted cluster-cluster interaction potential (dotted red
line). }\end{figure}

The fitted potential is displayed in \Rfg{potfit}(a). When applied to
the model system shown in \Rfg{plosdd}, it reproduces the long-range
peak of the distance distribution (see \Rfg{potfit}(b)). We also
checked if a similar agreement can be reached by varying only the
persistence length of the central ssDNA link. With larger stiffness,
this peak appears always shifted to the right and its maximum
approaches the experimental position only for very small $l_b$ of
ssDNA. In the absence of the cluster-cluster repulsion, however, the
distribution simultaneously  acquires the shape of the dotted plot in
\Rfg{plosdd} and never resembles the experimental data.

As we see, the gold cluster SAXSI data can be accounted for
without hypothetical long-range stretching correlations, by taking
into account the exclusion volume of the reporter labels and assuming
a moderate long-range repulsion between them. The origin of this
repulsion is most probably electrostatic. The thioglucose shells are
nominally neutral, however, the clusters have a negative charge
detectable in electrophoresis \cite{Mathew-Fenn:08b}, suggesting that
the pKa of the glucose hydroxyl groups is reduced due to OH\rar S-Au
substitution and glucose-glucose interactions at the surface.
Additional repulsive forces should appear due to ion exclusion and
cluster polarization in the DNA field. A hole in a dielectric exposed
to external electric field can be viewed as a superposition of two
oppositely polarized substances, and two such holes interact via the
dipole-dipole potential. The same is true for filled holes with
dielectric permittivity different from that of the media
\cite{Bottcher:73}. Similar qualitative considerations suggest that
additional polarization equivalent to permanent Coulomb charges should
result from exclusion of condensed ions from the space occupied by the
clusters.

Passivated gold clusters have complex electron structures and their
dielectric properties are not known. A very high dielectric
permittivity can result from delocalized electrons in the nanocrystal
core, but low values are also possible if the cluster stability is due
to the closed superatom electron shells similar to noble gases
\cite{Walter:08}. A rough evaluation of the possible cluster
orientations near DNA shows that in both cases the induced
polarization interactions should be repulsive. For an
order-of-magnitude comparison we used the solution of the
Poisson-Boltzmann equation for the electrostatic field around a
charged sphere in an aqueous monovalent salt. The corresponding
cluster-cluster interaction potential is written as (Ref.
\onlinecite{Cantor:80a}, Ch.  22)
$$
\Phi(r)=\left(q^2/\varepsilon r\right)exp[\kappa(a-r)]/(1+\kappa a)
$$
where $r$ is the distance, $\varepsilon=80$ is the water dielectric
constant, $a$ is the cluster radius (1.7 nm), and $\kappa$ is the
Debye-H\"uckel screening parameter (0.33 nm$^{-1}$ in 10 mM salt).
This formula gives a reasonably good agreement with the above fitted
potential if the cluster charge $q$ is about 12e (\Rfg{potfit}(a), the
dashed black line).  A complete quantitative analysis of this complex
system requires additional studies including accurate treatment of the
finite particle size, multipole interactions, and electrostatic
saturation. Nevertheless, \Rfg{potfit}(a) demonstrates that the
experimental data are reproduced with reasonable interaction energies.


Our interpretation suggests that, in close proximity to the DNA
surface, the electrostatic screening along its axis is not described
by the Debye-H\"uckel theory. The experimental distance variances are similar
under 10 mM, 100mM and 1M NaCl \cite{Mathew-Fenn:08b}, which was
interpreted as the absence of electrostatic interactions, but in fact
proves only the absence of strong salt dependence in both the bending
DNA stiffness and the cluster-cluster interactions. The first
assertion agrees with some earlier experimental data
\cite{Hagerman:88,Baumann:97}.  The environment of the double helix
strongly differs from the average solution conditions, notably, the
local ion concentration near DNA is higher than 1M regardless of the
net amount of the added salt. The cluster-cluster interactions occur
through DNA and the condensed ion layer, which does not correspond to
the classical Debye-H\"uckel screening. The long-range repulsion might
not change significantly with the NaCl concentrations used in
experiments \cite{Mathew-Fenn:08a,Mathew-Fenn:08b}. In agreement with
this assumption, the interaction potential fitted to the data
collected in 100 mM NaCl (\Rfg{scidd}) appears applicable to the
distance distribution obtained in 1M NaCl (\Rfg{potfit}(b)).

The foregoing analysis and the proposed alternative interpretation of
the experimental data involve strong assumptions that require further
verification. However, these assumptions are physically reasonable and
they do not contradict the available data, in contrast to the original
hypothesis of accordion stretching correlations in the DNA double
helix \cite{Mzlanl:09}. If this mechanism is confirmed, the gold
cluster SAXSI would present a promising tool for probing electric
fields near DNA and other polyelectrolyte surfaces. Effective
particle-particle interactions due to solvent polarization around the
double helix are important, for instance, for association of
multi-protein complexes on DNA. Additional insights in these
interactions can be obtained by direct atomistic MD simulations.  This
work is now in progress.

\end{document}